\documentclass[twocolumn,aps,superscriptaddress,showpacs]{revtex4}

\usepackage{amssymb}
\usepackage{amsmath}
\usepackage{graphicx}
\usepackage[normalem]{ulem}
\usepackage[dvips]{color}
\usepackage{multirow}
\usepackage{appendix}

\makeatletter

\newcommand{\Rmnum}[1]{\expandafter\@slowromancap\romannumeral #1@}
\makeatother

\begin{document}

\title{Simulating spin dynamics with spin-dependent cross sections in heavy-ion collisions}
\author{Yin Xia}
\affiliation{Shanghai Institute of Applied Physics, Chinese Academy
of Sciences, Shanghai 201800, China}
\affiliation{University of Chinese Academy of Sciences, Beijing 100049, China}
\author{Jun Xu\footnote{corresponding author: xujun@sinap.ac.cn}}
\affiliation{Shanghai Institute of Applied Physics, Chinese Academy
of Sciences, Shanghai 201800, China}
\author{Bao-An Li}
\affiliation{Department of Physics and Astronomy, Texas A$\&$M
University-Commerce, Commerce, TX 75429-3011, USA}
\affiliation{Department of Applied Physics, Xi'an Jiao Tong University, Xi'an 710049, China}
\author{Wen-Qing Shen}
\affiliation{Shanghai Institute of Applied Physics, Chinese Academy
of Sciences, Shanghai 201800, China}
\date{\today}

\begin{abstract}
We have incorporated the spin-dependent nucleon-nucleon cross
sections into a Boltzmann-Uehling-Uhlenbeck transport model for the
first time, using the spin-singlet and spin-triplet nucleon-nucleon
elastic scattering cross sections extracted from the phase-shift analyses of nucleon-nucleon
scatterings in free space. We found that the spin splitting of the
collective flows is not affected by the spin-dependent cross
sections, justifying it as a good probe of the in-medium nuclear
spin-orbit interaction. With the in-medium nuclear spin-orbit
mean-field potential that leads to local spin polarization, we found
that the spin-averaged observables, such as elliptic flows of free
nucleons and light clusters, becomes smaller with the spin-dependent
differential nucleon-nucleon scattering cross sections.

\end{abstract}


\maketitle

\section{Introduction}
\label{introduction}

The spin-orbit interaction, which was previously introduced to
explain the magic number of finite nuclei, is critical in
understanding the structures of rare isotopes and their impacts on
astrophysics~\cite{Rei95,Fur98,Ben99,Gau06,Sch13}. Heavy-ion
collisions provide the only way of studying properties of nuclear
matter as well as nuclear interactions at both finite densities and
temperatures in terrestrial laboratories, and a useful means of
extracting properties of the in-medium nuclear spin-orbit
interaction with optimal reaction conditions. Recently, we have
developed a spin- and isospin-dependent Boltzmann-Uehling-Uhlenbeck
(SIBUU) transport model, by incorporating the nucleon spin degree of
freedom and the nuclear spin-orbit interaction into the IBUU
transport model~\cite{Xu13,Xia216}. We found this model is useful in
studying the spin dynamics in intermediate-energy heavy-ion
collisions~\cite{Xu15}. In particular, it was observed that the spin
splittings of collective flows of free nucleons and light clusters
can be good probes of the in-medium spin-orbit
interaction~\cite{Xia14,Xia116}. However, in our previous studies,
we applied the spin-dependent mean-field potential for nucleons but
employed the spin-averaged nucleon-nucleon scattering cross
sections. In order to have a complete framework of the
spin-dependent transport approach and a better description of the
spin dynamics in intermediate-energy heavy-ion collisions, in the
present study we incorporated the spin-singlet and spin-triplet
cross sections for elastic nucleon-nucleon scatterings into the
model. The latter are extracted based on the phase-shift analyses of
nucleon-nucleon scatterings in free space. We found that the spin
splitting of the collective flow as a probe of the in-medium nuclear
spin-orbit interaction is almost not affected by the spin-dependent
nucleon-nucleon scattering cross sections. However, the overall
elliptic flows of free nucleons and light clusters are slightly
smaller with the spin-dependent nucleon-nucleon scattering cross
sections compared with the spin-averaged ones, if there is local
spin polarization induced by the spin-dependent mean-field
potential. A more complete BUU framework including both the
spin-dependent potential and the spin-dependent cross sections has
been established, providing possibilities of further exploring the
interesting physics of spin dynamics in intermediate-energy
heavy-ion collisions.

\section{Spin-dependent cross sections from phase-shift analyses}
\label{formulas}

The phase-shift analysis has been an effective way of decoupling
nucleon-nucleon interactions into various channels by fitting
experimental nucleon-nucleon scattering data in terms of the
scattering matrix~\cite{Win47,Wol51,Hau52,Mac59}. There are series
of studies on the energy-dependent phase-shift analyses of
nucleon-nucleon elastic scattering data in a wide energy
range~\cite{Arn77,Arn83,Arn87}. Using the phase-shift data in
Ref.~\cite{Arn77}, we evaluate the spin-singlet and spin-triplet
nucleon-nucleon elastic cross sections within the incident nucleon
energy range between 1 and 500 MeV, where the inelastic scatterings
are less important. For completeness, we first recall the most
relevant formulaes in the following.

We begin with the general formula for the differential cross section
of two-body collisions expressed directly in terms of the
eigenphases of the scattering matrix by Blatt and
Biedenharn~\cite{Bla52}:
\begin{equation}
d\sigma^{{\alpha}'{s}';\alpha s}=
\frac{g}{(2s+1)k^{2}}\sum_{L=0}^{\infty}B_{L}({\alpha}'{s}';\alpha s)P_{L}(\cos\theta)d\Omega ,\label{dsi}
\end{equation}
where $g$ is 1 for neutron-proton scatterings and 4 for
proton-proton (neutron-neutron) scatterings, $P_{L}(\cos\theta)$ is
the Legendre polynomial, $k$ is the center-of-mass (C.M.) momentum
in the two-body system, and the coefficients
\begin{eqnarray}
B_{L}({\alpha}'{s}';\alpha s)
&=&\frac{(-)^{{s}'-s}}{4}\sum_{J_{1}}\sum_{J_{2}}\sum_{l_{1}}\sum_{l_{2}}\sum_{{l_{1}}'}\sum_{{l_{2}}'} \nonumber\\
&\times& Z(l_{1}J_{1}l_{2}J_{2},sL)Z({l_{1}}'J_{1}{l_{2}}'J_{2},{s}'L) \nonumber\\
&\times& R.P.[(\delta _{{\alpha}'\alpha}\delta_{{s}'s}\delta  _{{l_{1}}'l_{1}}-S_{{\alpha}'{s}'{l_{1}}';\alpha s l_{1}}^{J_{1}} )^{*} \nonumber\\
&\times& (\delta _{{\alpha}'\alpha}\delta_{{s}'s}\delta _{{l_{2}}'l_{2}}-S_{{\alpha}'{s}'{l_{2}}';\alpha s l_{2}} ^{J_{2}})] \label{ble}
\end{eqnarray}
can be determined directly from the phase-shift data. In the above
expression, $\delta_{ab}$ represents the Kronecker $\delta$
function; $\alpha$, $s$, $l$, and $J$ represent the scattering
channel, the spin of the scattering channel, the orbital angular
momentum, and the total angular momentum, respectively;
$S_{{\alpha}'{s}'{l}';\alpha s l}^{J}$ is the scattering amplitude
of a collision from a channel $\alpha s l$ to a channel
${\alpha}'{s}'{l}'$; the $Z$ coefficient represents the selection
rules introduced by Biedenharn {\it et al.}~\cite{BBR52}; the $R.P.
[...]$ represents the real part of the expression in the bracket. In
the limit of pure elastic nucleon-nucleon scatterings without spin
flipping, ${\alpha}'=\alpha$ and ${s}'=s$ are always satisfied, so
we omit the superscript $\alpha$ and use only $s$ as the superscript
in the following. $s=0$ and $s=1$ stand for the spin-singlet and
spin-triplet scattering, respectively.

Let's first consider the spin-singlet and spin-triplet channel for
neutron-proton scatterings. For the spin-singlet case with $s=0$,
there is only one channel $l=J$. Using $S=\exp(2i\delta_{J}^0)$,
Eq.~(\ref{ble}) becomes
\begin{eqnarray}
B_{L}(0; 0)
&=&\sum_{J_{1}}\sum_{J_{2}}\sum_{l_{1}=J_{1}}\sum_{l_{2}=J_{2}}
{Z(l_{1}J_{1}l_{2}J_{2},0L)}^2       \nonumber\\
&\times& \sin \delta^0_{J_{1}} \sin \delta^0_{J_{2}} \cos (\delta^0_{J_{1}}-\delta^0_{J_{2}}),
\end{eqnarray}
where $\delta_J^0$ is the phase-shift of the spin-singlet channel
with orbital angular momentum $l=J$. For the spin-triplet case with
$s=1$, given $l=J$, there is still only one channel with
$S=\exp(2i\delta^1_{J})$. When the neutron-proton scattering is
affected by the tensor force in their spin-triplet state, the
angular momentum $l$ can have two values, i.e., $l=J\pm1$. In the
latter case, the general expression of the $S$ matrix of a
two-channel reaction is

\begin{scriptsize}
\begin{eqnarray}
&&S=\\
&&\hspace{-0.5cm}\begin{pmatrix}
{\cos}^{2}(\epsilon_J)  e^{2i\delta^1 _{J-1}}+{\sin}^{2}(\epsilon_J)  e^{2i\delta^1 _{J+1}} & \frac{1}{2}\sin(2\epsilon_J)(e^{2i\delta^1 _{J-1}}-e^{2i\delta^1 _{J+1}})\\
\frac{1}{2}\sin(2\epsilon_J)(e^{2i\delta^1 _{J-1}}-e^{2i\delta^1 _{J+1}}) & {\sin}^{2}(\epsilon_J)  e^{2i\delta^1 _{J-1}}+{\cos}^{2}(\epsilon_J)  e^{2i\delta^1 _{J+1}}
\end{pmatrix}.\notag \label{sm}
\end{eqnarray}
\end{scriptsize}

\noindent In the above, $\delta^1_{J\pm1}$ is usually called the
Biedenharn-Blatt (BB) phase shift of the spin-triplet channel with
$l=J\pm1$, and $\epsilon_J$ is the parameter describing the mixing
probability of the two coupling states. By using the
energy-dependent neutron-proton phase-shift data as well as the
mixing parameters for various channels in Tables \Rmnum{3} and
\Rmnum{4} of Ref.~\cite{Arn77}, we calculate the
coefficient $B_L$ and the differential cross section. For the
unpolarized neutron-proton cross section, we also take the summation
of the isovector contribution $T=1$, the isoscalar contribution
$T=0$, and their interference contribution to the coefficient
$B_L$~\cite{Xiath}. We note there is a simplified method for calculating the spin-triplet case developed by Blatt and
Biedenharn~\cite{Bie52}, and it leads to identical results.

For proton-proton scatterings, we only incorporate the nuclear
contribution to the cross section into transport model simulations, but subtract the contribution of the long-range Coulomb potential to the scatterings. For the spin-singlet and spin-triplet proton-proton
scatterings with $l=J$, the scattering matrix $S$ can be expressed
as
\begin{eqnarray}
S = e^{2i \delta^{0(1)}_{J}}-e^{2i \phi_{J}}+1,  \label{sl}
\end{eqnarray}
where $\phi_{J}$ is the pure Coulomb phase shift for orbital angular
momentum $l=J$, and it can be written as~\cite{Mot49}
\begin{equation}
\phi_{l}=\sum_{m=1}^{l}\arctan(\eta/m), \label{cou}
\end{equation}
with $\eta=e^{2}/\hbar v\approx {(137\beta)}^{-1}$ where $\beta=v/c$
is the reduced velocity of the incident proton in the lab frame.
In order to subtract the Coulomb contribution from the $S$ matrix
for the two channels of spin-triplet scatterings with $l=J \pm 1$,
we express it as~\cite{Sta57}

\begin{scriptsize}
\begin{eqnarray}
&&S=1+ \\
&&\begin{pmatrix}
\cos(2\epsilon_{J})  e^{2i\delta^1 _{J-1}}-e^{2i \phi_{J-1}} & i\sin(2\epsilon_{J}) e^{i(\delta^1 _{J+1}+\delta^1 _{J-1})}\\
i\sin(2\epsilon_{J}) e^{i(\delta^1 _{J+1}+\delta^1 _{J-1})} & \cos(2\epsilon_{J})  e^{2i\delta^1 _{J+1}}-e^{2i \phi_{J+1}}
\end{pmatrix}, \notag\label{smb}
\end{eqnarray}
\end{scriptsize}

\noindent with
\begin{eqnarray}
\delta^{0(1)}_{l} = \delta^{0(1)}_{l}(N)+\phi_{l}
\end{eqnarray}
where $\delta^{0(1)}_{l}(N)$ is called the nuclear bar phase shifts,
which are taken from Table \Rmnum{2} of Ref.~\cite{Arn77} for
various proton incident energies. The way of subtracting the Coulomb
contribution assumes that the Coulomb force acts only outside the
region of the nuclear force where the WKB approximation is
valid~\cite{Sta57}. For the spin-singlet and spin-triplet
proton-proton scatterings with $l=J$, the expressions for the
scattering matrix $S$ are the same for BB phase shifts and bar phase
shifts, as can be seen from Eq.~(\ref{sl}). In this way the
spin-dependent differential proton-proton elastic scattering cross
sections can also be obtained.

\section{Parametrization of the spin-dependent cross sections}
\label{parameterization}

Starting from the energy-dependent phase-shift results of
nucleon-nucleon scatterings by Arndt {\it et al.}~\cite{Arn77}, and
using the method described above, we are now able to obtain the
differential cross sections for elastic nucleon-nucleon scatterings
at various collision energies. Since the higher-order terms of the Legendre
polynomials vanish after intergration, the total cross section is determined by the terms
with $L=0$ in Eq.~(\ref{dsi}). The total cross sections for both
spin-singlet and spin-triplet elastic neutron-proton scatterings
between 1 and 500 MeV can be parameterized respectively as

\begin{footnotesize}
\begin{eqnarray}
\sigma_{np}^{0}&=&9302.64/E^{3}-982.187/E^{2}-1.32\times10^{2}+1.03E\notag \\
&-&3.06\times10^{-3}E^{2}+4.85\times10^{-6}E^{3}-3.19\times10^{-9}E^{4},\\
\sigma_{np}^{1}&=&-27888.84\times10^{4}/E^{3}+17565.39/E^{2}+ 13382.81/E\notag \\&-&8.10\times10^{1}+0.37E-2.75\times10^{-4}E^{2}-9.60\times10^{-7}E^{3} \notag \\
&+&1.37\times10^{-9}E^{4}.
\end{eqnarray}
\end{footnotesize}

\noindent Similarly, the total cross sections for both spin-singlet
and spin-triplet elastic proton-proton scatterings between 1 and 500
MeV can also be parameterized respectively as

\begin{footnotesize}
\begin{eqnarray}
\sigma_{pp}^{0}&=&-11877.31/E^{3}+733.31/E^{2}+ 17397.66/E-2.38\times10^{2}\notag \\&+&1.51E-4.9\times10^{-3}E^{2}+8.37\times10^{-6}E^{3}-5.58\times10^{-9}E^{4} , \notag \\
\sigma_{pp}^{1}&=&-1.20+0.79E-8.40\times10^{-3}E^{2}+3.24\times10^{-5}E^{3},\notag \\
 && ( 1~\text{MeV}< E<100~\text{MeV}) \notag \\
\sigma_{pp}^{1}&=&1.72\times10^{1}+0.16E-8.13\times10^{-4}E^{2}+2.14\times10^{-6}E^{3}      \notag \\
&-&2.86\times10^{-9}E^{4}+1.52\times10^{-12}E^{5}.   \notag \\
 && (100~\text{MeV} < E<500~\text{MeV})
\end{eqnarray}
\end{footnotesize}

\begin{figure}[h]
\includegraphics[scale=0.3]{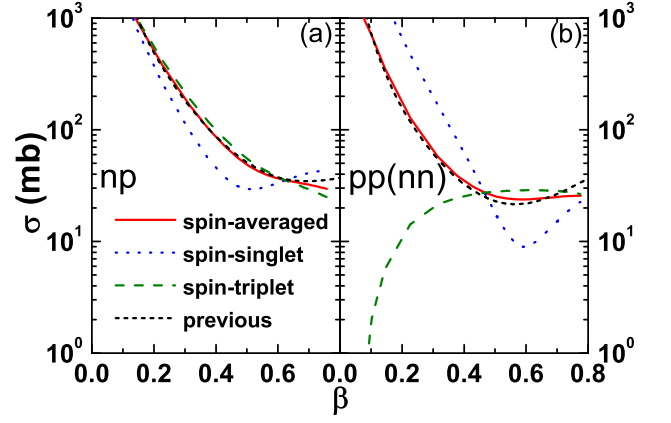}
\caption {(color online) Total elastic spin-averaged, spin-singlet,
and spin-triplet cross sections obtained in the present study for
neutron-proton (left) and proton-proton (neutron-neutron) (right) scatterings as
functions of the reduced incident nucleon velocity $\beta$ in the
lab frame, compared with the previous ones used in the IBUU
transport model.}\label{tcs}
\end{figure}

\noindent In the above, $\sigma^0$ and $\sigma^1$ in mb are the
cross sections for the elatistic spin-singlet and spin-triplet
scatterings, respectively, and $E$ in MeV is the kinetic energy of
the incident nucleon in the lab frame. The spin-averaged cross
section can be obtained from $\sigma
=\sigma^{0}/4+3\sigma^{1}/4$. In Fig.~\ref{tcs} we compared the total
elastic scattering cross sections obtained in the present study
with those previously used in the IBUU transport model, with the
latter taken from Ref.~\cite{Cha90} parameterized as
\begin{eqnarray}
\sigma_{pp(nn)}=8.76/\beta^{2}-15.04/\beta+13.73+68.76\beta^{4},\\
\sigma_{np}=25.26/\beta^{2}-18.18/\beta-70.67+113.85\beta,
\end{eqnarray}
where the cross section $\sigma$ is in mb, and
$\beta=\sqrt{1-M^{2}_{N}c^4/(M_{N}c^2+E)^{2}}$ is the reduced velocity of
the incident nucleon with $M_{N}$ being the nucleon mass. We find
that the previously used parameterized cross sections are similar to
the spin-averaged ones obtained in the present study using the
phase-shift results in the energy range considered. Note that a cut at
very low energy region, where the cross section may diverge, is
usually used in transport model simulations.

We have also parameterized the differential spin-singlet and
spin-triplet cross sections for elastic neutron-proton and
proton-proton scatterings between 1 and 500 MeV in the following
form:
\begin{eqnarray}
d\sigma^{s}_{np}(\theta) &=& {\sum_{n=0}^{11}} a^{s}_{n}{\cos}^{n}\theta d\Omega ,\\
d\sigma^{s}_{pp}(\theta) &=& {\sum_{m=0}^{5}} b^{s}_{2m}{\cos}^{2m}\theta d\Omega .
\end{eqnarray}
In the above equations, the cross sections are in mb, $n$ and $m$
are related to the angular momentum quantum numbers of the orbital
wave function, i.e., from $s$-wave to $h$-wave, as used in the
energy-dependent phase-shift analyses. With the differential cross
sections from phase-shift results at discrete energies, we are able
to parameterize the coefficients $a^s_{n}$ and $b^s_{2m}$ as
functions of the energy $E$ to get continuous energy-dependent
differential cross sections between 1 and 500 MeV. The $s$-wave
coefficients $a^0_{0}$, $a^1_{0}$, $b^0_{0}$, and $b^1_{0}$, which
lead to the total cross section, are parameterized respectively as

\begin{footnotesize}
\begin{eqnarray}
a^0_{0}&=&-47.99/E^{3}-28.93/E^{2}+740.42/E-12.11+7.53\times10^{-2}E\notag \\
&-&2.32\times10^{-4}E^{2}+3.83\times10^{-7}E^{3}-2.62\times10^{-10}E^{4},\\
a^1_{0}&=&-2435.74/E^{3}+1565.07/E^{2}+1115.4/E-9.27+0.02E \notag\\
&+&1.43\times10^{-4}E^{2}-6.04\times10^{-7}E^{3}+6.08\times10^{-10}E^{4}, \\
b^0_{0}&=&-987.99/E^{3}+80.497/E^{2}+ 1409.47/E-23.51+0.148E \notag\\
&-&4.488\times10^{-4}E^{2}+7.203\times10^{-7}E^{3}-4.72\times10^{-10}E^{4}, \\
b^1_{0}&=&-0.142+0.079E-8.78\times10^{-4}E^{2}+3.36\times10^{-6}E^{3}, \notag \\
&& (1~\text{MeV} <E< 100~\text{MeV}) \notag \\
b^1_{0}&=&1.89+9.85\times10^{3}E-6.90\times10^{-5}E^{2}+1.84\times10^{-7}E^{3}       \\
&-&2.33\times10^{-10}E^{4}+1.16\times10^{-13}E^{5} .   \notag \\
&& (100~\text{MeV} < E < 500~\text{MeV}) \notag
\end{eqnarray}
\end{footnotesize}

\noindent For other coefficients $a^s_{n}$ and $b^s_{2m}$
corresponding to larger orbital angular momentum quantum numbers,
polynomial functions are used to fit their energy dependence, and
the fitting results are showed in Table~\ref{T1}.

\begin{figure}[h]
    \includegraphics[scale=0.3]{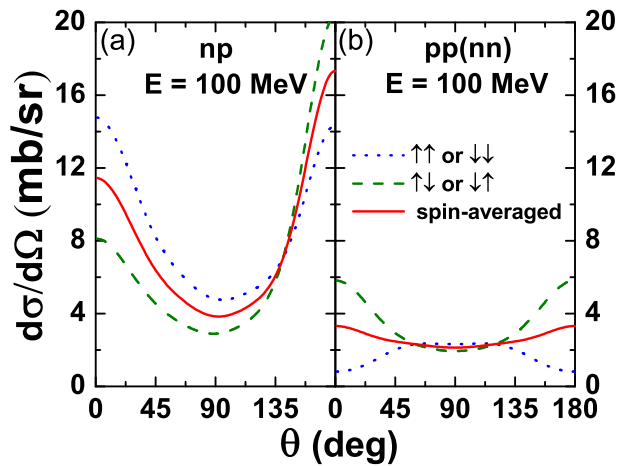}
    \caption {(color online) Differential cross sections for neutron-proton (left)
    and proton-proton (right) pairs with the same spin, different spins, and the
    spin-averaged differential cross sections as functions of the scattering polar
    angle $\theta$ in the C.M. frame, at an incident nucleon energy of 100 MeV.}\label{dcs}
\end{figure}

In transport model simulations of heavy-ion collisions, it is more
convenient to use the cross sections determined by the spins of the
colliding nucleons, and they can be expressed in terms of the
spin-singlet and spin-triplet scattering cross section as
\begin{eqnarray}
\sigma^{\uparrow\uparrow(\downarrow\downarrow)}_{NN} &=& \sigma^1_{NN}, \\
\sigma^{\uparrow\downarrow(\uparrow\downarrow)}_{NN}&=&(\sigma^1_{NN}+\sigma^0_{NN})/2,
\end{eqnarray}
where $\sigma^{\uparrow\uparrow(\downarrow\downarrow)}_{NN}$
($\sigma^{\uparrow\downarrow(\uparrow\downarrow)}_{NN}$) is the
cross section for nucleon pairs with the same (different) spin with
respect to the angular momentum of the pair. The angular dependence
of the differential cross sections to be used in SIBUU transport
model simulations is plotted in Fig.~\ref{dcs}. These angular
distributions reveal the nucleon interaction in vacuum. For example,
in neutron-proton scatterings, the forward peak is due to the Wigner
force while the backward peak is due to the Majorana
force~\cite{book}. On the other hand, scatterings between identical
particles with the same spin and isospin are not likely to have
forward and backward peaks due to the strong Pauli repulsive effect.

\begin{table*}[htbp]\footnotesize
    \caption{Coefficients for the polynomial fit of the energy dependence of $a^s_{n}$ $(n>0)$ and $b^s_{2m}$ $(m>0)$, i.e., $a^s_{n}(b^s_{2m})=C_{0}+C_{1}E+C_{2}E^{2}+C_{3}E^{3}~(1~\text{MeV}<E<50~\text{MeV})$ and $a^s_{n}(b^s_{2m})=C_{4}+C_{5}E+C_{6}E^{2}+C_{7}E^{3}+C_{8}E^{4}+C_{9}E^{5}~(50~\text{MeV} < E < 500~\text{MeV})$.   }
 \begin{tabular}{|c|cccc|cccccc|}
    \hline
    & $C_{0}$ & $C_{1}$ &  $C_{2}(10^{-3})$ & $C_{3}(10^{-5})$ &  $C_{4}$ &  $C_{5}$ &  $C_{6}(10^{-3})$ &  $C_{7}(10^{-6})$ & $C_{8}(10^{-9})$ &  $C_{9}(10^{-12})$  \\
    \hline
    $a^0_{1}$  & $-6.23$ & $-0.46$ & $32.24$ & $-42.98$ & $-2.20$ & $-2.84\times10^{-3}$ & $0.14$ & $-0.26$ &$-0.064$ & $0.29$ \\
    $a^0_{2}$  & $-0.48$ & $0.54$ & $-21.99$ & $24.79$ & $4.71$ & $-7.55\times10^{-2}$ & $0.74$ & $-2.41$ &$3.54$ & $-1.98$ \\
    $a^0_{3}$  & $0.47$ & $-0.32$ & $7.92$ & $-4.77$ & $-8.82$ & $0.23$ & $-1.98$ & $6.80$ &$-10.61$ & $6.39$ \\
    $a^0_{4}$  & $-4.07\times10^{-2}$ & $11.08$ & $5.23$ & $-9.12$ & $0.23$ & $0.12$ & $-1.96$ & $7.48$ &$-12.09$ & $7.36$ \\
    $a^0_{5}$  & $-0.13$ & $9.77\times10^{-2}$ & $-11.76$ & $15.66$ & $-1.80$ & $-0.12$ & $1.37$ & $-5.28$ &$8.92$ & $-5.75$ \\
    $a^0_{6}$  & $4.80\times10^{-2}$ & $-3.77\times10^{-2}$ & $4.57$ & $-5.47$ & $3.41$ & $-0.14$ & $3.03$ & $-12.81$ &$21.96$ & $-13.90$ \\
    $a^0_{7}$  & $-2.63\times10^{-2}$ & $1.90\times10^{-2}$ & $-1.99$ & $2.07$ & $1.43$ & $-0.11$ & $1.13$ & $-4.02$ &$6.57$ & $-4.01$ \\
    $a^0_{8}$  & $2.12\times10^{-2}$ & $-1.55\times10^{-2}$ & $1.68$ & $-2.23$ & $0.10$ & $0.12$ & $-2.67$ & $11.89$ &$-21.24$ & $13.77$ \\
    $a^0_{9}$  & $-2.52\times10^{-3}$ & $1.49\times10^{-3}$ & $-0.052$ & $-0.90$ & $2.43$ & $-6.06\times10^{-2}$ & $-0.30$ & $1.97$ &$-3.80$ & $2.55$ \\
    $a^0_{10}$  & $-4.21\times10^{-5}$ & $4.64\times10^{-4}$ & $-0.19$ & $1.45$ & $-4.37$ & $9.84\times10^{-2}$ & $0.43$ & $-3.68$ &$7.86$ & $-5.56$ \\
    \hline
    $a^1_{1}$  & $0.36$ & $-8.13\times10^{-2}$ & $7.28$ & $-10.15$ & $2.72$ & $-1.39\times10^{-2}$ & $-0.11$ & $0.92$ &$-2.11$ & $1.59$ \\
    $a^1_{2}$  & $-0.18$ & $-0.22$ & $16.87$ & $-21.24$ & $1.10$ & $1.07\times10^{-1}$ & $-1.00$ & $4.27$ &$-8.11$ & $5.63$ \\
    $a^1_{3}$  & $1.17\times10^{-2}$ & $1.15\times10^{-2}$ & $-1.34$ & $2.27$ & $-1.69$ & $5.93\times10^{-2}$ & $-0.57$ & $2.10$ &$-3.42$ & $2.09$ \\
    $a^1_{4}$  & $5.37\times10^{-2}$ & $-3.46\times10^{-2}$ & $3.80$ & $-6.46$ & $4.16$ & $-6.24\times10^{-2}$ & $-0.65$ & $4.21$ &$-8.40$ & $5.75$ \\
    $a^1_{5}$  & $-1.81\times10^{-3}$ & $3.65\times10^{-3}$ & $-1.07$ & $1.97$ & $-2.72$ & $6.95\times10^{-2}$ & $-0.36$ & $0.39$ &$0.41$ & $-0.72$ \\
    $a^1_{6}$  & $2.03\times10^{-2}$ & $-1.74\times10^{-2}$ & $2.24$ & $-0.78$ & $-5.71$ & $0.25$ & $-1.42$ & $3.63$ &$-4.75$ & $2.52$ \\
    $a^1_{7}$  & $-1.05\times10^{-2}$ & $8.01\times10^{-3}$ & $-0.91$ & $1.06$ & $1.76$ & $-0.11$ & $1.41$ & $-5.04$ &$7.87$ & $-4.64$ \\
    $a^1_{8}$  & $1.32\times10^{-3}$ & $-9.45\times10^{-4}$ & $0.09$ & $0.08$ & $-0.70$ & $2.77\times10^{-2}$ & $-0.19$ & $0.53$ &$-0.66$ & $0.29$ \\
    $a^1_{9}$  & $-3.39\times10^{-4}$ & $5.50\times10^{-5}$ & $0.05$ & $-0.61$ & $0.94$ & $-1.36\times10^{-2}$ & $-0.43$ & $2.03$ &$-3.58$ & $2.27$ \\
    $a^1_{10}$  & $-5.08\times10^{-5}$ & $6.50\times10^{-5}$ & $-0.01$ & $0.09$ & $-0.08$ & $-1.16\times10^{-3}$ & $0.11$ & $-0.48$ &$0.81$ & $-0.48$ \\
    \hline
    \hline
    $b^0_{2}$  & $-1.24$ & $-9.83\times10^{-1}$ & $-36.74$ & $38.86$ & $8.51$ & $-9.91\times10^{-2}$ & $0.47$ & $-1.27$ &$1.84$ & $-1.04$ \\
    $b^0_{4}$  & $0.16$ & $-0.12$ & $14.36$ & $-20.55$ & $4.01$ & $4.33\times10^{-2}$ & $-0.87$ & $3.07$ &$-4.73$ & $2.91$ \\
    $b^0_{6}$  & $1.44\times10^{-2}$ & $-1.02\times10^{-2}$ & $0.99$ & $-0.35$ & $-1.62$ & $0.064$ & $0.0061$ & $-0.27$ &$0.57$ & $-0.61$ \\
    $b^0_{8}$  & $1.38\times10^{-4}$ & $-8.66\times10^{-5}$ & $0.0048$ & $0.034$ & $-0.078$ & $0.0022$ & $0.0097$ & $-0.042$ &$0.068$ & $-0.041$ \\
    \hline
    $b^1_{2}$  & $0.039$ & $-0.019$ & $-0.82$ & $2.06$ & $-1.44$ & $0.03$ & $-0.15$ & $0.52$ &$-0.98$ & $0.67$ \\
    $b^1_{4}$  & $-0.027$ & $0.023$ & $-1.79$ & $1.63$ & $-0.488$ & $0.397$ & $-1.61$ & $3.08$ &$-2.10$ & $5.63$ \\
    $b^1_{6}$  & $6.53\times10^{-5}$ & $-3.52\times10^{-3}$ & $0.52$ & $-0.84$ & $2.13$ & $-5.18\times10^{-2}$ & $0.34$ & $-0.84$ &$0.65$ & $-0.073$ \\
    $b^1_{8}$  & $1.73\times10^{-3}$ & $-1.95\times10^{-3}$ & $0.204$ & $-0.18$ & $-0.63$ & $4.94\times10^{-2}$ & $-0.68$ & $2.55$ &$-3.93$ & $2.25$ \\
    $b^1_{10}$  & $-2.30\times10^{-3}$ & $6.16\times10^{-3}$ & $-0.021$ & $0.15$ & $-0.29$ & $-3.78\times10^{-3}$ & $0.23$ & $-1.04$ &$1.75$ & $-1.07$ \\
    \hline
    \end{tabular}
    \label{T1}
\end{table*}

\section{Effects in heavy-ion simulations}
\label{results}

In the SIBUU transport model, the density of the initial two nuclei
is sampled according to the prediction of Skyrme-Hartree-Fock
calculations, while the momentum of each nucleon is sampled
according to its local density and further boosted by the beam
energy. The spin expectation value of each nucleon is chosen as a
unit vector in the $4\pi$ solid angle, and it is randomly sampled in
the initial stage. As the system begins to evolve, the coordinate
$\vec{r}$, momentum $\vec{p}$, and spin $\vec{s}$ of each nucleon
follow the equations of motion consistently derived from the
spin-dependent Boltzmann-Vlasov equation~\cite{Xia216} as follows:
\begin{eqnarray}
d\vec{r}/dt &=& \vec{p}/\sqrt{p^2c^2+M_N^2c^4} + \nabla_p U^{s}, \\
d\vec{p}/dt &=& -\nabla U - \nabla U^{s}, \\
d\vec{s}/dt &=& -\frac{i}{\hbar} [\vec{s}, U^s],
\end{eqnarray}
where $U$ is the momentum- and spin-independent mean-field
potential, and the right-hand side of the third equation denotes the
commutator of each component of spin with the spin-dependent
mean-field potential $U^s$. Particularly, the strength, the isospin
dependence, and the density dependence of $U^s$ are still under
debate and are hot topics in nuclear structure studies~\cite{Xu15}.
In our previous studies, we have shown that nucleons with different
spins may have different dynamics with $U^s$, and this leads to
local spin polarization (see Fig.~1 of Ref.~\cite{Xu13} and Fig.~1
of Ref.~\cite{Xia116}) as well as the spin splitting of collective
flows of free nucleons and light clusters~\cite{Xu13,Xia14,Xia116}.
Here we investigate the effects of nucleon-nucleon scatterings with
spin-dependent differential elastic cross sections in heavy-ion
collisions. Since the spin expectation direction is known for each nucleon, the spin state of a single nucleon and that of the colliding nucleon pair can be obtained by projecting the spin expectation direction onto the total angular momentum of the incoming nucleon pair. The differential scattering cross sections are then determined from the spin state as well as the collision energy through the parameterizations given in Sec.~\ref{parameterization}, and they are technically incorporated according to the
scattering treatment in Appendix B of Ref.~\cite{Ber88}.

\begin{figure}[h]
\includegraphics[scale=0.3]{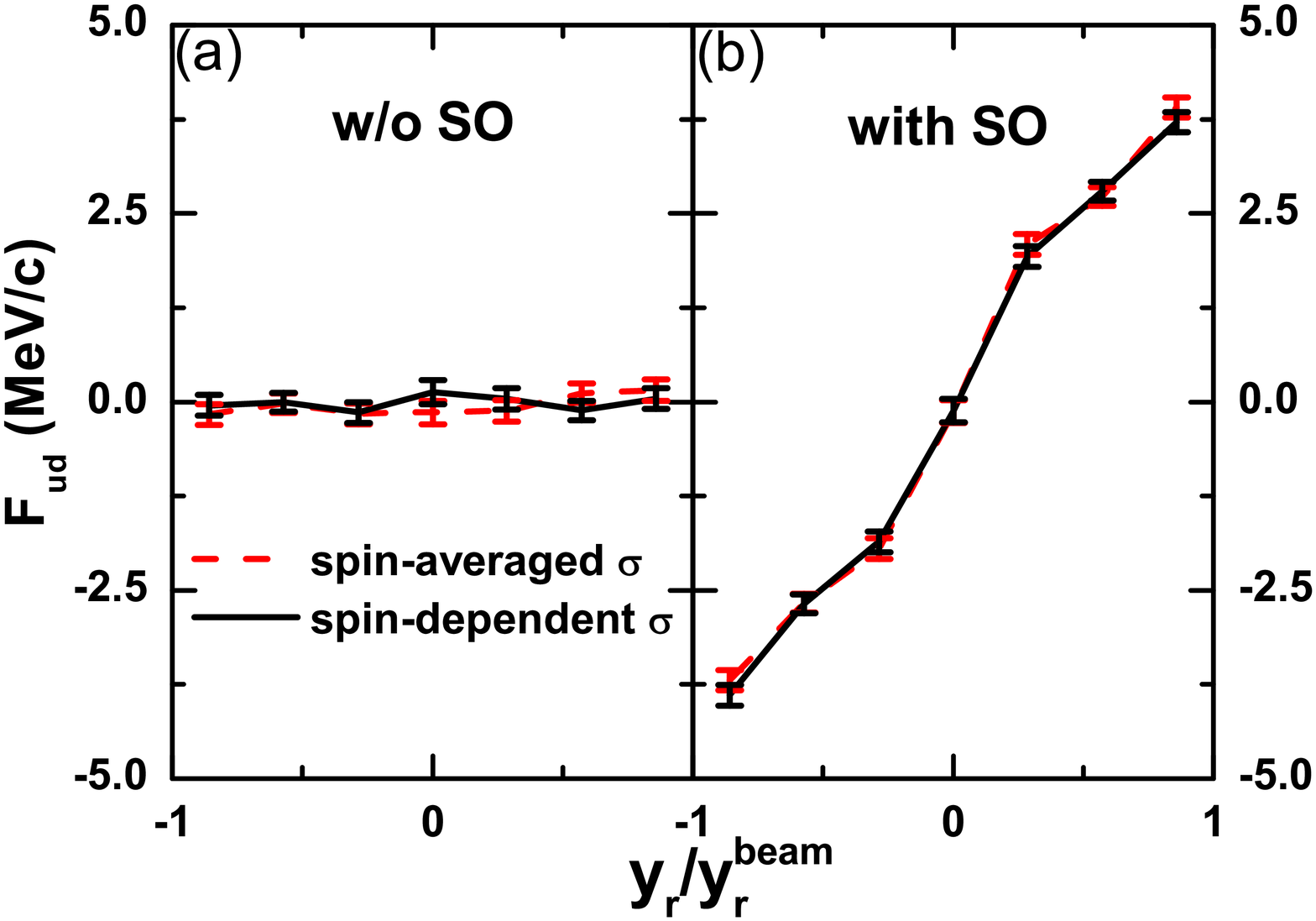}
\caption {(color online) Rapidity dependence of the spin differential transverse flow from the spin-averaged
and the spin-dependent cross sections with (right) and without
(left) the spin-orbit (SO) potential in Au+Au collisions at 100
MeV/nucleon and an impact parameter $\text{b}=12$ fm. }\label{Fud}
\end{figure}

We first examine the effects of the spin-dependent cross sections on
the spin up-down differential transverse flow defined
as~\cite{Xu13,Xia14}
\begin{equation}
F_{ud}(y_r) = \frac{1}{N(y_r)} \sum_{i=1}^{N(y_r)} s_i (p_x)_i,
\end{equation}
where $N(y_r)$ is the number of nucleons with rapidity $y_r$,
$(p_x)_i$ is the momentum of the $i$th nucleon in $x$ direction, and
$s_i$ is $1 (-1)$ for spin-up (spin-down) nucleons with respect to
the total angular momentum of the heavy-ion collision system. As
discussed in Refs.~\cite{Xu13,Xia14}, the spin-dependent potential
$U^s$ gives an additional attractive (repulsive) potential to
spin-up (spin-down) nucleons, resulting in their different
transverse flows. As shown in the left panel of Fig.~\ref{Fud},
$F_{ud}$ vanishes without $U^s$, with the latter the source of
different potentials for spin-up and spin-down nucleons and thus
their different transverse flows. With $U^s$, $F_{ud}$ remain almost
the same using the spin-averaged and spin-dependent nucleon-nucleon
cross sections, as shown in the right panel of Fig.~\ref{Fud}. We
have also found that the spin up-down differential transverse flow
remains the same for neutrons and protons as well as for energetic
nucleons, justifying the validity of $F_{ud}$ as a good probe of the
strength, the isospin dependence, and the density dependence of the
in-medium nuclear spin-orbit potential~\cite{Xu13,Xia14,Xia116}.

\begin{figure}[h]
\includegraphics[scale=0.3]{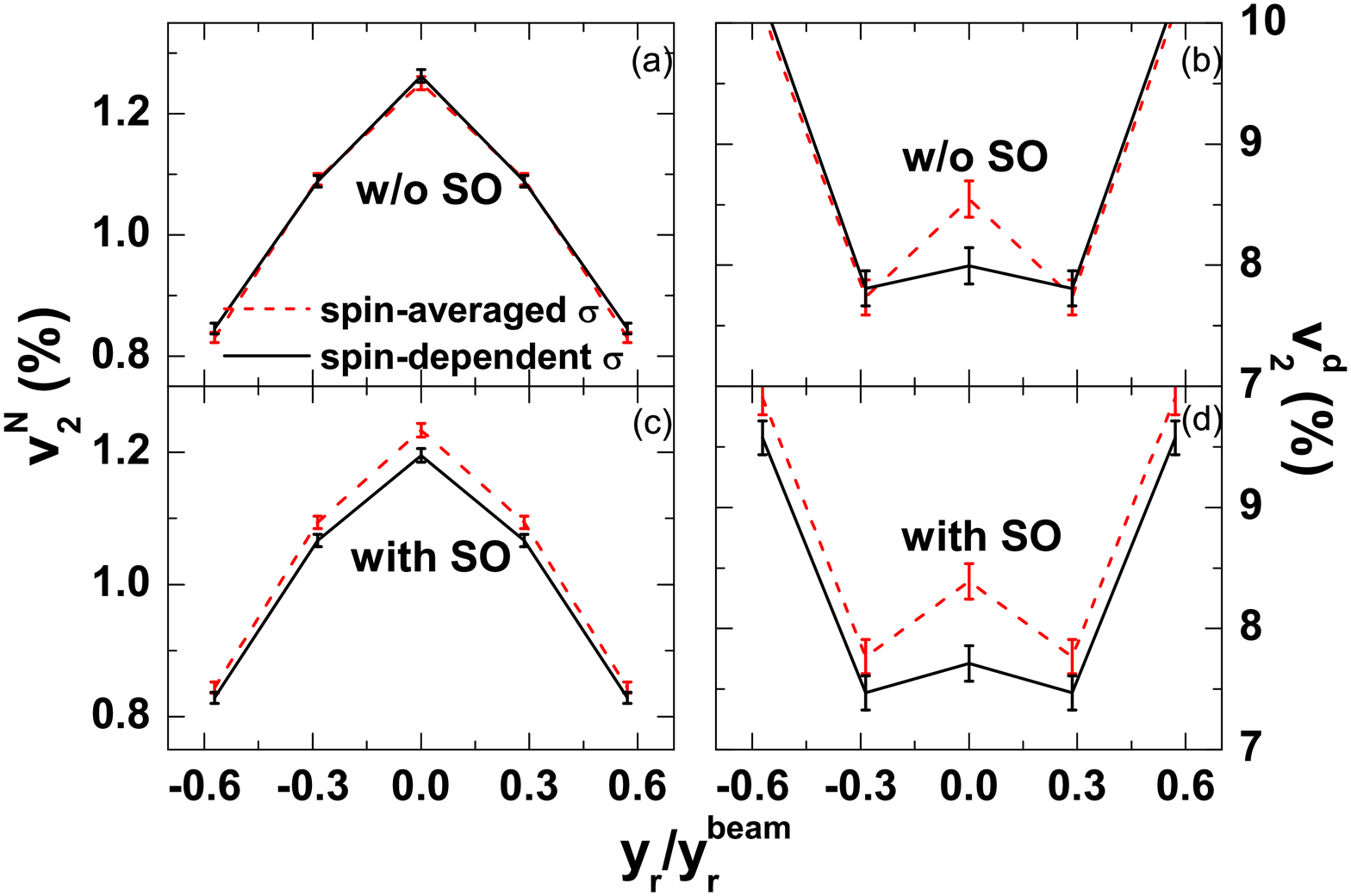}
\caption {(color online) Rapidity dependence of the elliptic flow of
free nucleons (left) and deuterons (right) from the spin-averaged
and the spin-dependent cross sections with (lower) and without
(upper) the spin-orbit (SO) potential in Au+Au collisions at 100
MeV/nucleon and an impact parameter $\text{b}=12$ fm. }\label{v2}
\end{figure}

Figure~\ref{v2} compares the resulting spin-averaged elliptic flows
($v_2=\langle (p_x^2-p_y^2) / (p_x^2+p_y^2) \rangle$) of free
nucleons and deuterons from the spin-averaged and spin-dependent
nucleon-nucleon scattering cross sections with and without the
spin-dependent mean-field potential. It is seen that $v_2$ of free
nucleons is the same from the spin-averaged and spin-dependent cross
sections without $U^s$, while the difference is observed with $U^s$.
The latter is due to the local spin polarization induced by $U^s$.
As is know, $v_2$ is sensitive to the shear viscosity of the
system~\cite{Son11,Zho14}, with the later related to the transport
cross section defined as
\begin{equation}
\sigma_{tr} = \int \frac{d\sigma}{d\Omega} (1-\cos^2\theta).
\end{equation}
For a given total $\sigma$, a more forward- and backward-peaked
differential cross section generally leads to a smaller transport
cross section and a larger shear viscosity. Since locally there can
be different numbers of spin-up and spin-down nucleons induced by
the spin-dependent mean-field potential, the transport cross section
can be different from the spin-averaged and spin-dependent cross
sections. This is the reason why the slightly different $v_2$ is
observed in Panel (c) of Fig.~\ref{v2}. We have also studied the
formation of light clusters formed in transport simulations, through
a dynamical coalescence algorithm from nucleons that are close in
coordinate and momentum space~\cite{Chen03,Xia116}. The
spin-dependent differential cross sections lead to correlations
between the scattering angles of final-state nucleons, resulting in
the different final distribution of these light clusters. It is seen
from Fig.~\ref{v2} that slight $v_2$ difference is observed at
midrapidities between results from the spin-averaged cross sections
and the spin-dependent ones, and the effect is further enhanced with
the spin-dependent mean-field potential.

\section{Conclusion and outlook}
\label{summary}

Using the phase-shift results from nucleon-nucleon scattering data
in free space, we have obtained the spin-dependent neutron-proton
and proton-proton differential elastic scattering cross sections. We
have further incorporated them into the spin-dependent
Boltzmann-Uehling-Uhlenbeck transport model for the first time. The
spin splittings of collective flows, which were previously found as
probes of the in-medium nuclear spin-orbit interaction, are not
affected by these spin-dependent cross sections. However,
spin-averaged quantities, such as the elliptic flows of free
nucleons and deuterons, can be slightly affected with both the
spin-dependent mean-field potential and cross sections.

We note that it is still a big challenge to obtain the
spin-dependent inelastic nucleon-nucleon scattering cross sections
in the suitable energy range for intermediate-energy heavy-ion
collisions, due to the lack of the experimental data. On the other
hand, the in-medium nucleon-nucleon scattering cross sections remain uncertain, even for the spin-independent part. So far, the information of the in-medium cross sections relies on various many-body theories~\cite{Li94,Li00,Zha07}, while in transport model simulations the in-medium effective mass scaling~\cite{Pan91,Li05} or empirical parameterizations~\cite{Kla93} are generally used. We are still on the way of looking for reliable probes of the
in-medium nuclear spin-orbit interaction and studying interesting
and relevant physics of spin dynamics in heavy-ion collisions, by
using a dynamical framework as complete as possible.

\begin{acknowledgments}
We thank Xiao-Ming Xu for helpful communications, and Chen Zhong for
maintaining the high-quality performance of the computer facility.
This work was supported by the Major State Basic Research
Development Program (973 Program) of China under Contract Nos.
2015CB856904 and 2014CB845401, the National Natural Science
Foundation of China under Grant Nos. 11320101004, 11475243, and
11421505, the "100-talent plan" of Shanghai Institute of Applied
Physics under Grant Nos. Y290061011 and Y526011011 from the Chinese
Academy of Sciences, the Shanghai Key Laboratory of Particle Physics
and Cosmology under Grant No. 15DZ2272100, the U.S. Department of Energy,
Office of Science, under Award Number de-sc0013702, and the CUSTIPEN
(China-U.S. Theory Institute for Physics with Exotic Nuclei) under
the US Department of Energy Grant No. DEFG02- 13ER42025.
\end{acknowledgments}


\begin{thebibliography}{99}

\bibitem{Rei95} P. G. Reinhard and H. Flocard, Nucl. Phys. A {\bf 584}, 467 (1995).

\bibitem{Fur98} R. J. Furnstahl, John J. Rusnak, and Brian D. Serot, Nucl. Phys. A {\bf 632}, 607 (1998).

\bibitem{Ben99} M. Bender {\it et al.}, Phys. Rev. C {\bf 60}, 034304 (1999).

\bibitem{Sch13} J. P. Schiffer {\it et al.}, Phys. Rev. Lett. {\bf 110}, 169901 (2013).

\bibitem{Gau06} L. Gaudefroy {\it et al.}, Phys. Rev. Lett. {\bf 97}, 092501 (2006).

\bibitem{Xu13} J. Xu and B. A. Li, Phys. Lett. B {\bf 724}, 346 (2013).

\bibitem{Xia216}  Y. Xia, J. Xu, B. A. Li, and W. Q. Shen, Phys. Lett. B {\bf 759}, 596 (2016).

\bibitem{Xu15} J. Xu, B. A. Li, W. Q. Shen, and Y. Xia, Front. Phys. {\bf 10}, 102501 (2015).

\bibitem{Xia14} Y. Xia, J. Xu, B. A. Li, and W. Q. Shen, Phys. Rev. C {\bf 89}, 064606 (2014).

\bibitem{Xia116}  Y. Xia, J. Xu, B. A. Li, and W. Q. Shen, Nucl. Phys. A {\bf 955}, 41 (2016).

\bibitem{Win47} E. P. Winger and L. Eisenbud, Phys. Rev. {\bf 72}, 29 (1947).

\bibitem{Wol51} L. Wolfenstein, Phys. Rev. {\bf 82}, 690 (1951).

\bibitem{Hau52} W. Hauser and H. Feshbach, Phys. Rev. {\bf 87}, 366 (1952).

\bibitem{Mac59} M. H. Macgregor, Phys. Rev. {\bf 113}, 1559 (1959).

\bibitem{Arn77} R. A. Arndt {\it et al.}, Phys. Rev. C {\bf15}, 1002 (1977).

\bibitem{Arn83} R. A. Arndt {\it et al.}, Phys. Rev. D {\bf28}, 97 (1983).

\bibitem{Arn87} R. A. Arndt {\it et al.}, Phys. Rev. D {\bf35}, 128 (1987).

\bibitem{Bla52} J. M. Blatt and L. C. Biedenharn, Rev. Mod. Phys. {\bf24}, 258 (1952).

\bibitem{BBR52} L. C. Biedenharn, J. M. Blatt, and M. E. Rose, Rev. Mod. Phys. {\bf24}, 249 (1952).

\bibitem{Xiath} Y. Xia, Ph.D thesis, Shanghai Institute of Applied Physics, 2017.

\bibitem{Bie52} J. M. Blatt and L. C. Biedenharn, Phys. Rev. {\bf 86}, 399 (1952).

\bibitem{Mot49} N. F. Mott and H. S. W. Massey, {\it Theory of Atomic Collisions} (Clarendon Press, Oxford, 1949), second edition, p. 101.

\bibitem{Sta57} H. P. Stapp, T. J. Ypsilantis, and N. Metropolis, Phys. Rev. {\bf 105}, 302 (1957).

\bibitem{Cha90} S. K. Charagi and S. K. Gupta, Phys. Rev. C {\bf41}, 1610 (1990).

\bibitem{book} J. M. Blatt and V. F. Weisskopf, {\it Theoretical
Nuclear Physics} (Springer-Verlag, New York Inc., 1979), p. 178.

\bibitem{Ber88} G. F. Bertsch and S. Das Gupta, Phys. Rep.  {\bf 160}, 189 (1988).

\bibitem{Son11} H. C. Song, S. A. Bass, U. Heinz, T. Hirano, and C. Shen, Phys. Rev. Lett. {\bf 106}, 192301 (2011).

\bibitem{Zho14} C. L. Zhou, Y. G. Ma, D. Q. Fang, G. Q. Zhang, J.
Xu, X. G. Cao, and W. Q. Shen, Phys. Rev. C {\bf 90}, 057601 (2014).

\bibitem{Chen03} L. W. Chen, C. M. Ko, and B. A. Li, Nucl. Phys. A {\bf 729}, 809 (2003).

\bibitem{Li94} G. Q. Li and R. Machleidt, Phys. Rev. C {\bf 49}, 566 (1994).

\bibitem{Li00} Q. F. Li, Z. X. Li, and G. J. Mao, Phys. Rev. C {\bf 62}, 014606 (2000).

\bibitem{Zha07} H. F. Zhang, Z. H. Li, U. Lombardo, P. Y. Luo, F. Sammarruca, and W. Zuo, Phys. Rev. C {\bf 76}, 054001 (2007).

\bibitem{Pan91} V. R. Pandharipande and S. C. Pieper, Phys. Rev. C {\bf 45}, 791 (1991).

\bibitem{Li05} B. A. Li and L. W. Chen, Phys. Rev. C {\bf 72}, 064611 (2005).

\bibitem{Kla93} D. Klakow, G. Welke, and W. Bauer, Phys. Rev. C {\bf 48}, 1982 (1993).

\end{thebibliography}
\end{document}